# Selective Excitation of Vibrations in a Single Molecule

Yang Luo[1,*], Shaoxiang Sheng[1], Michele Pisarra[2,3], Alberto Martin-Jimenez[1,4], Fernando Martin[4,5,*], Klaus Kern[1,6], Manish Garg[1,*]

[1] Max Planck Institute for Solid State Research, Heisenbergstr. 1, 70569 Stuttgart, Germany
[2] Dipartimento di Fisica, Università della Calabria, Via P. Bucci, cubo 30C, 87036, Rende (CS), Italy
[3] INFN-LNF, Gruppo Collegato di Cosenza, Via P. Bucci, cubo 31C, 87036, Rende (CS), Italy
[4] Instituto Madrileño de Estudios Avanzados en Nanociencia (IMDEA Nano), Faraday 9, Cantoblanco, 28049 Madrid, Spain
[5] Departamento de Química, Módulo 13, Universidad Autónoma de Madrid, 28049 Madrid, Spain
[6] Institut de Physique, Ecole Polytechnique Fédérale de Lausanne, 1015 Lausanne, Switzerland

*Authors to whom correspondence should be addressed.
y.luo@fkf.mpg.de, fernando.martin@imdea.org and mgarg@fkf.mpg.de

**Abstract**

**The capability to excite, probe, and manipulate vibrational modes is essential for understanding and controlling chemical reactions at the molecular level. Recent advancements in tip-enhanced Raman spectroscopies have enabled the probing of vibrational fingerprints in a single molecule with Ångström-scale spatial resolution. However, achieving controllable excitation of specific vibrational modes in individual molecules remains challenging. Here, we demonstrate the selective excitation and probing of vibrational modes in single deprotonated phthalocyanine molecules utilizing resonance Raman spectroscopy in a scanning tunneling microscope. Selective excitation is achieved by finely tuning the excitation wavelength of the laser to be resonant with the vibronic transitions between the molecular ground electronic state and the vibrational levels in the excited electronic state, resulting in the state-selective enhancement of the resonance Raman signal. Our approach sets the stage for steering chemical transformations in molecules on surfaces by selective excitation of molecular vibrations.**

**Introduction**

The vibrational modes of a molecule play a crucial role in its chemical (and geometrical) transformation[1,2] as well as in many energy conversion phenomena[3-5]. Various theoretical and experimental studies have demonstrated the pivotal role of selectively exciting molecular vibrations in achieving precise control over the desired output of chemical transformations and electron transport processes[6-8]. To comprehend the elementary mechanisms of molecular reactions, much effort has been directed toward exciting vibrational modes and modifying chemical reactions at the single-molecule level[9-11]. Scanning tunneling microscopy (STM) has emerged as a powerful technique to investigate single-molecule events related to specific vibrational modes in real-space, utilizing inelastic electron tunneling scattering[12-14]. Nevertheless, the broad energy distribution of tunneling electrons will inevitably populate the excited states of the molecule within a wide energy range, thus making it difficult to unequivocally probe one specific vibrational mode.

The selective excitation of vibrational modes in molecules is usually achieved by resonant infrared laser radiation[15,16] or by stimulated Raman pumping[17,18]. However, the combination of infrared spectroscopy with scanning probe microscopy can hardly achieve sub-nanometer resolution[19], which is crucial for single-molecule studies. The recent advancements in integrating STM with visible and near-infrared light have expanded the capability of manipulating vibrational degrees of freedom at the single-molecule level[20-23]. STM based tip-enhanced Raman spectroscopy (TERS)[24-28] has demonstrated Ångström-scale resolution in probing the spatial distribution of vibrational modes in a single molecule. Nevertheless, this enhanced

Raman spectroscopy involves inelastic scattering from numerous vibration modes instead of specific (selective) modes.

A promising approach to enhance particular molecular vibrations is through resonance Raman scattering, which relies on the coupling between vibrational and electronic states in molecules[29-34]. By tuning the excitation wavelength to fall within the optical absorption band of the molecules, specific vibrational modes coupled to an excited electronic state can be selectively excited. This technique has been successfully applied to reveal structural changes and functional dynamics of photoactive chromophores in the bulk phase[35-37]. Moreover, by aligning the excitation photon energy to different vibronic transitions, resonant Raman scattering can be used to precisely determine the electronic and vibrational structure of molecules in the bulk phase[38,39], and to study the exciton-phonon interactions in low-dimensional materials[40-42].

In this work, we employ tip-enhanced resonance Raman scattering to investigate the mode selective excitation in a single molecule. Resonance Raman scattering and fluorescence signals were simultaneously measured from deprotonated phthalocyanine molecules ($HPc^-$) adsorbed on a thin insulating film on top of a metallic substrate. The contributions of the Raman scattering and fluorescence processes are identified by their emission peak positions and linewidths in the recorded spectra. We demonstrate selective excitation of the vibrational modes by tuning the excitation wavelength of the laser to be resonant with the corresponding molecular vibronic transitions between the ground electronic state and specific vibrational modes in the excited electronic state. The mode selective excitation manifests as a strong enhancement of particular vibrational modes (associated with the chosen vibronic transitions) in the recorded resonance Raman spectra. To elucidate the physical mechanism underlying the mode selective excitation, we performed first-principles density functional theory (DFT) based simulations accounting for the Raman transitions involving different modes in the excited and ground electronic states of the molecule, which agree well with the experimental findings. Furthermore, we performed spatial dependence measurements of the resonance Raman scattering to illustrate the Ångström-scale resolution in selective excitation of molecular vibrations.

**Results**

**Single-molecule spectroscopy**

In our experiments, we used STM-based near-field spectroscopic techniques including tip-enhanced photoluminescence (TEPL)[43-46] and STM-induced electroluminescence (STML)[47-51] to characterize the optical spectral properties of single free base phthalocyanine ($H_2Pc$) molecules and their deprotonated species ($HPc^-$), as illustrated in Fig. 1a. A gold tip and an Au(111) substrate were used to achieve a strong plasmonic enhancement for the used laser wavelength (see Methods). An ultrathin three-monolayer (3ML) sodium chloride (NaCl) insulating film was deposited on the Au(111) surface to electronically decouple the single molecules from the underlying Au substrate, which efficiently suppresses the quenching rates of the excited molecular excitons.

Figure 1b shows the STML spectrum of an $H_2Pc$ molecule (blue curve in the top panel of Fig. 1b) locally excited by tunneling electrons at the bias voltage of 2.4 V. A main emission peak at ~1.815 eV with a full width at half maximum (FWHM) of ~8 meV was measured, which is assigned to a purely electronic contribution, usually referred to as $Q_x$, in the phthalocyanine molecules[52-55]. The molecular TEPL spectrum was measured by using a continuous wave (CW) laser centered at ~633 nm at a bias voltage of −1.0 V, as shown by the red curve in the top panel of Fig. 1b. Here, the energy of tunneling electrons (up to 1 eV) is much lower compared to the electronic transition energy (~1.8 eV) of the $H_2Pc$ molecule, which is not sufficient to excite the $H_2Pc$ molecule directly as in the STML measurement[56]. In contrast to the STML spectrum showing a single broad fluorescence peak, the TEPL spectrum of the $H_2Pc$ molecule shows additional distinct narrow peaks over a broad peak.

A deprotonated version of the $H_2Pc$ molecule was also investigated by STML and TEPL measurements[53,57]. The deprotonation of an $H_2Pc$ molecule is achieved by positioning the STM tip on top of the molecular center and applying a bias voltage of ~ 3 V. The STML spectrum of the $HPc^-$ molecule shows a blue-shifted fluorescence peak at ~1.86 eV (blue curve in the bottom panel of Fig. 1b) compared to the $H_2Pc$, possibly induced by the internal Stark effect in the deprotonated molecule[53]. The TEPL spectrum of the $HPc^-$ molecule shows sharp peaks sitting on top of a broad peak, similar to the spectral features measured from the $H_2Pc$ molecule.

The broad peaks in the TEPL spectra for both types of molecules can be attributed to the fluorescence emission, whose linewidth is determined by the total electronic damping time, which is usually ~10 meV for single molecules in the STM junction[43]. The appearance of narrow emission peaks overlaying on top of the broad fluorescence background (Fig. 1b) implies the involvement of either the Raman scattering processes or the vibronic transitions between the vibrational levels of the involved electronic states, i.e., excited and ground electronic states, in the TEPL spectra.

**Selective excitation of molecular vibrations**

To decipher the origin of the narrowband emission lines appearing in the TEPL spectra, we tuned the excitation wavelength of the laser in the TEPL measurements. While tuning the laser's wavelength will not alter the vibronic transition energies as they are intrinsic molecular properties, it will shift the energies of Raman scattering light, which depend on the energy differences of the incident laser photon energy and the molecular vibrational modes[58]. For this experiment, we excited the molecule using a continuously wavelength-tunable picosecond fiber laser (see methods for details). We note that the $H_2Pc$ molecule is prone to deprotonation upon photoexcitation with the picosecond laser, possibly caused by the strong photo-induced electron tunneling on interaction with the ultrashort laser pulses. Thus, we focused on the $HPc^-$ molecule in the wavelength-dependent experiments.

The TEPL spectra of an $HPc^-$ molecule were recorded at various excitation wavelengths ranging from 631 nm to 641.5 nm as shown in Fig. 2a. We note that the picosecond laser pulses have a spectral linewidth of ~1.5 nm (~5 meV at 640 nm), which reduces the spectral resolution in the TEPL spectra as compared with the CW laser excitation (Fig. 1b). As a result, the bandwidth of the narrow emission peaks in Fig. 2a is limited to ~5 meV. An apparent gradual shift of the narrow emission peaks, as indicated by the colored arrows in Fig. 2a, can be seen upon decreasing the excitation wavelength of the laser, while the broad fluorescence peak remains at the same energy. As explained below, this indicates that the measured narrow emission peaks overlapping with the fluorescence peak have their origin in Raman scattering processes.

By converting the horizontal axis in Fig. 2a from photon energy (eV) to the relative Raman shift frequency ($cm^{-1}$) with respect to the laser excitation wavelength, we obtain the wavelength-dependent Raman excitation map, as shown in Fig. 2b. The spectrally invariable fluorescence peak appears to gradually shift with the excitation wavelength in the Raman shift representation. Notably, several narrow peaks at ~555, 615, 680, 725, and 790 $cm^{-1}$ (indicated by black arrows), with fixed Raman shift energies appearing on top of the broad fluorescence peak can be clearly identified. This constant energy shift with excitation wavelength clearly shows that the narrow emission peaks overlapping with the fluorescence emission have their origin in the Raman scattering processes. A striking feature in the Raman excitation map is that the observed Raman peaks undergo an intensity modulation upon the variation of the excitation wavelength of the laser.

To understand the underlying mechanism of the observed features in the experiment, we illustrate in Fig. 3a the energy diagrams of the fluorescence and Raman scattering processes (see Supplementary Section I and Supplementary Fig. 1 for details). In the fluorescence process, the molecule is photo-excited from its ground electronic state $|0, g\rangle$ to the vibrational levels in the excited electronic state $|v_e, e\rangle$. Following photoexcitation, the molecule can redistribute the excess vibrational energy among its various modes through an internal conversion process or exchange energy with its environment to reach the bottom of the excited electronic state, i.e., $|0, e\rangle$, whence fluorescence emission to the ground electronic state takes place. On the other hand, the Raman scattering process connects the initial $|0, g\rangle$ state and the final, $|v_g, g\rangle$ states, without involving a direct photon absorption in the molecule. Yet, the Raman signal can be dramatically enhanced when the excitation wavelength of the laser is resonant with the electronic absorption band, a process referred to as resonance Raman scattering[29,33].

Within the Born-Oppenheimer approximation, the resonant Raman scattering cross section is mainly determined by two vibronic transitions: (1) from $|0, g>$ to $|v_e, e>$, and (2) from $|v_e, e>$ to $|v_g, g>$. When the energy of the incident photons matches a vibronic transition, e. g., $|0, g>$ to $|v_e, e>$, the transition rate is maximized, leading to an increase in the Raman scattering cross-section. For instance, the resonant excitation of the $v_I$ vibrational mode leads to a strong resonant Raman signal from $v_I$, as shown in Fig. 3b (left panel), while resonant excitation of the $v_{II}$ vibrational mode at higher photon energy leads to an enhancement of Raman scattering from $v_{II}$ (right panel). In this way, the selective enhancement of Raman scattering from a particular vibrational mode can be achieved by tuning the laser photon energy to be resonant with its related vibronic transition. The intensity of the fluorescence and Raman lines are further modulated by the Franck-Condon (FC) overlap integrals involving the vibrational wave functions of the upper, $|v_e>$, and lower, $|v_g>$, electronic states, $<v_g|v_e>$. When the geometries of the molecule in the ground and excited electronic states are similar, the energy spacing of the vibrational levels in both electronic states are also very similar and the largest Franck-Condon overlaps correspond to transitions involving identical vibrational modes in the two electronic states (indicated by thick arrows in Fig. 3a and 3b). As a consequence, the energy difference of the dominant Raman transition will be nearly identical to that of the fluorescence transition, which is what we observe in our experiment (Fig. 2a).

Following the concept of resonance Raman scattering, we performed time-dependent DFT-based first-principles simulations (see Supplementary Section II for details) considering the Raman scattering between the vibronic levels of the ground and excited electronic states of the HPc⁻ molecule. For comparison with the experiment, the excited singlet electronic state with the lowest transition energy ($S_1$), corresponding to the $Q_x$ state as measured experimentally, was considered. The calculations include explicitly the dipole couplings between the ground (*g*) and the excited (*e*) electronic states and the Franck-Condon overlap integrals, $<v_g|v_e>$. The resonant Raman scattering spectra were obtained by calculating the Raman scattering cross-sections at various excitation energies. Fig. 3c shows the simulated Raman spectrum as a function of the energy detuning between the incident photons and the calculated optical gap between the lowest vibronic states $|0, g>$ and $|0, e>$. The simulated result matches quite well with the experimental Raman excitation map (Fig. 2b), which demonstrates the selective excitation of vibrational modes using resonant Raman scattering processes (Supplementary Fig. 5). This achievement stems from the precise control of the incident photon energy to align with a particular molecular vibronic transition rather than the $|0, g>$ to $|0, e>$ one. We note that non-selective resonant excitation of Raman modes in a single molecule has been earlier demonstrated by resonant excitation of this $|0, g>$ to $|0, e>$ transition, where all the Raman modes coupled to the excited state were enhanced[21]. It is worth emphasizing that the dephasing time of the vibronic states is sufficiently long in the experiments, which enables the spectral separation of different vibronic transitions and the observation of the dominance of specific Raman transitions from a single vibrational mode.

To obtain the energies of the molecular vibrational modes involved in the resonance Raman scattering processes, we summed up the TEPL spectra in the Raman excitation maps (Fig. 2b and Fig. 3c) to produce the integrated Raman spectrum of the HPc$^-$ molecule, as shown in Fig. 3d. The relative intensities of different vibrational modes in the generated spectrum is a consequence of the varying strengths of the Frank-Condon overlap integrals for different vibronic transitions over the tuned excitation wavelength. Here, we focus on the vibrational modes between 500 and 900 cm$^{-1}$ and the associated molecular vibronic transitions. Five distinct Raman modes can be distinguished in the simulated Raman spectrum (dashed-red curves in Fig. 3d), which is in good agreement with the measured spectrum (red curve in Fig. 3d), especially for the frequencies of the vibrational modes. The same Raman modes are observed in the calculated non-resonant Raman spectrum for the ground electronic state, as shown by the blue curve in Fig. 3d. The differences in the relative spectral intensities of the measured (red curve) and calculated (green curve) spectra might be attributed to the effect of the nanocavity plasmons on the molecular vibronic transition rates.

**Ångström-scale resolution in selective excitation of molecular vibrations**

To explore the extent of spatial confinement of selective excitation of molecular vibrations in the STM junction, TEPL spectra of a single HPc$^-$ molecule were recorded by increasing the plasmonic gap size, and by moving the nanotip laterally over the molecule (Fig. 4a and Fig. 4b). The molecule was excited by picosecond laser pulses set at ~633.5 nm. Fig. 4c shows a series of spectra measured by increasing the plasmonic gap size, i.e., by reducing the tunneling current under the constant current operating mode of the STM. The sharp peak at ~1.86 eV originates from the resonant Raman scattering for the vibrational mode present at ~790 cm$^{-1}$, and the broad peak results from the fluorescence emission. The intensities of resonance Raman scattering (black squares) and fluorescence (red dots) contributions in the measured spectra were obtained by integrating the spectral intensities over the area under the Raman and fluorescence peak positions, both of which decay exponentially with increasing plasmonic gap size, as shown in Fig. 4d. Fitting the intensity profiles with an exponential function (solid curves), $P \propto \exp(-\Delta z/k)$, yields decay constants of $k \sim 0.25$ nm for both of the contributions in the measured spectra, suggesting a very similar dependence of the resonance Raman and fluorescence signals on the plasmonic gap size. This is because the Raman scattering and fluorescence rates have the same dependence on the electronic transition dipolar moments between the ground and the excited electronic states. Such similarity arises from the same dependence of the electronic transition dipolar moments between $g$ and $e$ on both Raman scattering and fluorescence rates[30].

A series of TEPL spectra were recorded by positioning the nanotip along the blue crosses over the molecule, as shown in Fig. 4e. The integrated resonance Raman scattering (black squares) and fluorescence intensities (red dots) are plotted as a function of the lateral distance of the nanotip from the molecular center, as shown in Fig. 4f. The spatial confinement of the fluorescence emission shows a minimum at the molecular center, while it is maximum at the edge of the molecule. This emission behavior is determined by the dipolar interaction between the nanocavity plasmons and the molecule[43]. The integrated spectral intensities in Fig. 4e and 4f are plotted in absolute terms, implying that no TEPL signal is measured at the center of the molecule, pointing out the importance of achieving spatial resolution for an optimum vibrational selectivity. The spatial variation of the Raman signal shows similar feature with that of the fluorescence signal. From the evolution of the Raman scattering intensity change at various nanotip positions, we can estimate a spatial resolution of ~5 Ångström in the selective excitation of molecular vibrations in a single molecule.

**Conclusion**

In conclusion, we have demonstrated selective excitation of vibrations in a single molecule utilizing STM-based tip-enhanced resonance Raman scattering. By tuning the excitation wavelengths of the incident laser to resonance with the corresponding vibronic levels in the excited state, we can selectively excite and probe a specific vibrational mode in a single molecule with Ångström-scale resolution. Our approach serves as a unique tool to coherently control the population of selective vibrational modes and investigate mode-selectivity in photoexcitation of single molecules, which pave the way for engineering light-matter interactions at the sub-molecular level.

**Methods**

**Sample and tip preparation**

The experiments were performed in a custom-built scanning tunneling microscope (STM) operating in ultra-high vacuum conditions (~$2\times10^{-10}$ mbar), and at liquid helium temperature (~10 K). Au(111) surfaces were prepared by repeated cycles of sputtering with 1.0 keV $Ar^+$ ions and thermal annealing at ~500 °C. Au tips prepared by electrochemical etching were used in all the experiments to increase the electric field enhancement. $H_2Pc$ molecules were thermally sublimated onto the NaCl-covered Au(111) surface using a homemade evaporator with the substrate kept at liquid helium conditions (~10 K). All topographic images presented in this work were acquired in the constant current mode of the STM.

**Spectroscopic measurements**

The optical setup of the STM-based near-field spectroscopic measurements is shown in Supplementary Fig. 4. A Helium-Neon (He-Ne) CW laser (HNL150L, Thorlabs) centered at ~633 nm was used for the TEPL measurements. A supercontinuum white light fiber laser (SuperK FIANIUM) was used to generate the continuously wavelength-tuned laser beam. An achromatic lens (diameter: 50 mm; focusing length: 75 mm) was mounted inside the UHV chamber to focus the laser beams onto the apex of the Au tip. The molecular Raman and fluorescence signals were collected through the same achromatic lens and then focused into the entrance slit of a spectrometer (Kymera 328i, ANDOR) and detected by a thermoelectrically cooled charge-coupled device (iDus 416, ANDOR). A dichroic mirror centered at ~650 nm was used to separate the laser beam and the molecular signal.

**Resonance Raman Calculations**

The electronic structure DFT calculations have been performed with the Gaussian 16 package[59] using the B3LYP functional and the 631G(d,p) basis set. The resonance Raman scattering calculations have been carried out by fully taking into account the excited electronic state involved in the transition as implemented in the Gaussian 16 package[60,61] (see the Supplementary Section II for more details). First, a geometry optimization of the HPc$^-$ molecule in the electronic ground state was carried out and the vibrational frequencies and eigenvectors of the different normal modes in this electronic ground state were obtained. Then, the vertical electronic excitation spectrum and transition dipole moments at the ground state geometry were calculated within a Time Dependent-DFT (TDDFT) approach[62,63], obtaining up to the 20$^{th}$ electronic state (see table I in the SM). This procedure allowed us to identify the electronic transition from the $S_0$ to the $S_1$ states involved in the experiments. A geometry optimization was then performed for the molecule in the $S_1$ state, using as starting point the ground state geometry. For this configuration, a vibrational analysis was carried to obtain the vibrational energies and eigenvectors, which were plugged into the resonance Raman scattering calculation. In the Raman spectra, the energy of the vibrational modes was scaled by a factor of 0.96, determined by the comparison between the experimental Raman spectrum from $H_2Pc$ powder and the calculated Raman spectrum of the $H_2Pc$ molecule for the ground electronic state (Supplementary Fig. 6).

**Data availability**

The data that support the findings of this study are available from the corresponding authors on request.

**Code availability**

The DFT and TD-DFT simulations have been carried out using the Gaussian16 package[59]. The detailed information for the simulations are available from the corresponding authors on request.

## Acknowledgments

We thank Wolfgang Stiepany and Marko Memmler for technical support. This article is based upon work from the COST action CA18222 -- Attosecond Chemistry (AttoChem), supported by COST (European Cooperation in Science and Technology). All calculations were performed at the Centro de Computación Científica de la Universidad Autónoma de Madrid (CCC-UAM). MP acknowledges financial support by Centro Nazionale di Ricerca in High-Performance Computing, Big Data and Quantum Computing, PNRR 4 2 1.4, CI CN00000013, CUP H23C22000360005. FM acknowledges support by the Ministerio de Ciencia e Innovación MICINN (Spain) through projects PID2022-138288NB-C31 and the "Severo Ochoa" Programme for Centres of Excellence in R&D (CEX2020-001039-S). A.M-J acknowledges funding from HORIZON-MSCA-2022-PF-01-01 under the Marie Skłodowska-Curie grant agreement No. 101108851.

## Contributions

Y.L., S.S., M.G., A.M.J built the experimental set-up, performed the experiments and analyzed the experimental data. M.P. and F.M. designed and performed the theoretical calculations and analyzed the theoretical data. M.G. conceived the project and designed the experiments. K.K. supervised the project. All authors interpreted the results and contributed to the preparation of the manuscript.

## Competing interests

The authors declare no competing interests.

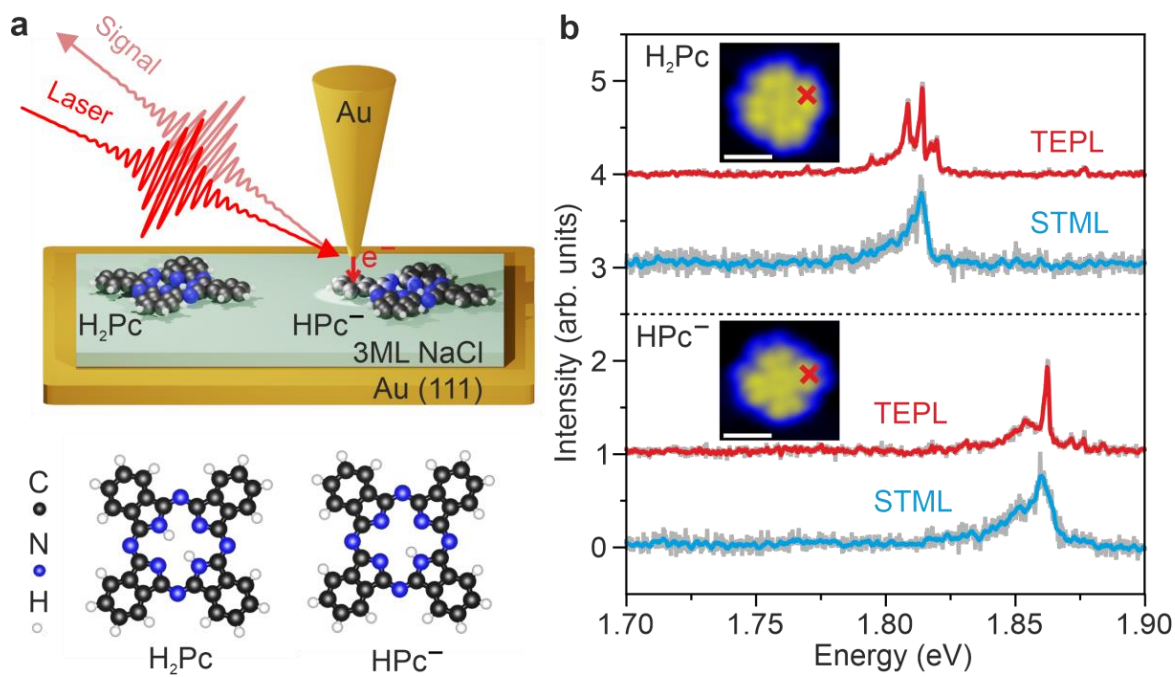

**Fig. 1: Single-molecule spectroscopy. a**, Top Panel: Schematic illustration of the experimental setup for STM-induced electroluminescence (STML) and tip-enhanced photoluminescence (TEPL) measured from single molecules adsorbed on three-monolayer thick NaCl on top of Au(111). Bottom panel: The structures of the $H_2Pc$ and $HPc^-$ molecules. **b**, STML (blue curves) and TEPL (red curves) spectra of individual $H_2Pc$ and $HPc^-$ molecules. The STML spectrum of $H_2Pc$ was recorded with tunneling electron excitation at a bias voltage $V = 2.4$ V, a tunneling current $I = 100$ pA with an acquisition time $t = 5$ s. The STML spectrum of $HPc^-$ is measured at $V = 2.4$ V, $I = 20$ pA, and $t = 60$ s. The TEPL spectra were recorded with CW laser excitation ($\lambda \sim 633$ nm and laser power of 0.2 mW) at $V = -1$ V, $I = 4$ pA, and $t = 10$ s. All spectra are normalized for clarity. The insets show the measured STM topography images of the $H_2Pc$ and $HPc^-$ molecules adsorbed on the three-monolayer thick NaCl/Au(111) surface, acquired at $V = -2$ V and $I = 4$ pA. The STML and TEPL spectra were recorded by placing the nanotip above the molecular lobe indicated by the red crosses in the topographic images. The white scale bars indicate a length of 1 nm.

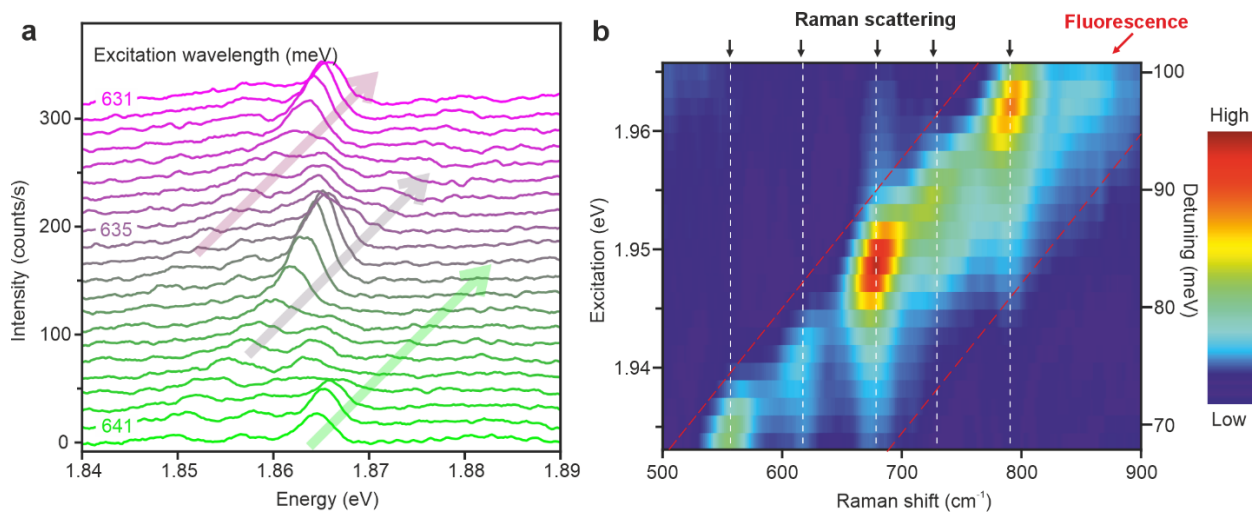

**Fig. 2: Wavelength-dependent photoexcitation of a single HPc¯ molecule. a**, A series of TEPL spectra of a single HPc¯ molecule recorded by tuning the excitation wavelength of the laser. The laser wavelength was tuned from 631 nm to 641.5 nm with a step size of 0.5 nm, and the laser power was kept constant at 0.23 mW. The TEPL spectra were measured by placing the nanotip above the molecular lobe (red-cross in Fig. 1b) at $V = -1.5$ V, $I = 4$ pA, and $t = 20$ s. The three arrows indicate the spectral shifting of the narrow emission peaks upon changing the excitation wavelength of the laser. The spectra are vertically shifted for clarity. **b**, Raman excitation map generated from the recorded TEPL spectra shown in **a**. The x-axis is the Raman shift frequency (cm$^{-1}$), and the left y-axis is the tunable laser excitation photon energy (eV). The right y-axis denotes the energy detuning (in meV) between the incident laser photon energy and the TEPL peak at ~1.865 eV. The black arrows on top of the figure and the vertical dashed-white lines indicate the spectral positions of the Raman peaks, whereas the red arrow and the space between the tilted dashed-red lines indicate the position of the broad fluorescence emission.

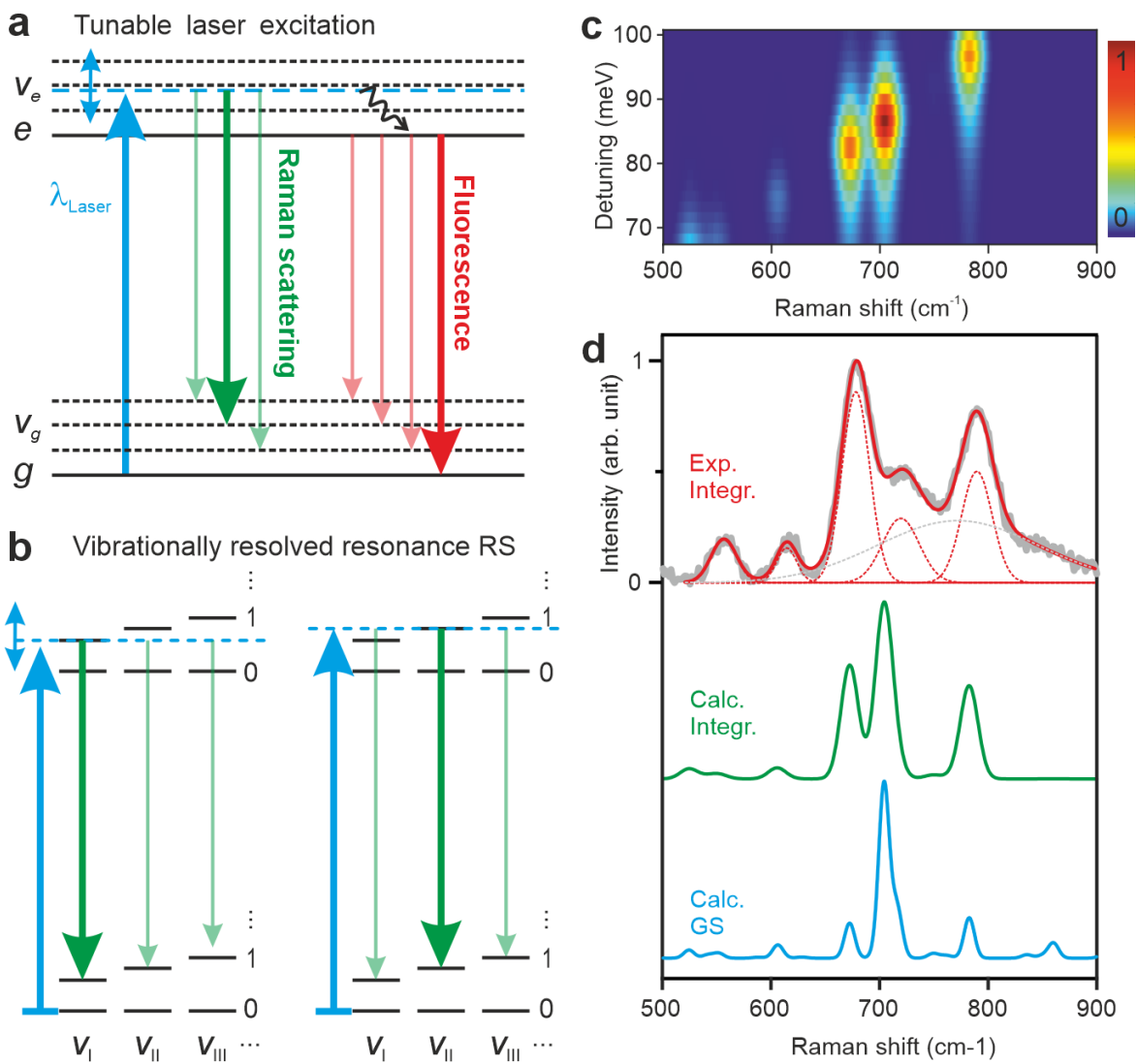

**Fig. 3: Vibrationally resolved resonance Raman spectroscopy of a single molecule. a**, Schematic illustration of the resonance Raman scattering and fluorescence in a molecule with tunable laser excitation. The Raman transitions are represented by green curves, whereas the fluorescence processes are indicated by the red curves. The relative intensities of different transitions are symbolized by the thickness of the downward arrows. **b**, Schematic illustration of the selective enhancement of Raman scattering from particular vibrational mode, $v_I$ or $v_{II}$, achieved by tuning the laser photon energy to be resonant with its related vibronic transition. Numerals, 0, 1… represent the quantum numbers of the various vibrational modes (e.g. $v_I$ or $v_{II}$) in the two electronic states. The molecule is initially in the vibrational ground state before the laser excitation. **c**, Simulated Raman excitation map generated by calculating the resonance Raman scattering spectra at various energy detunings between the incident photons and the calculated optical gap between the lowest vibrational states of the two electronic states $g$ and $e$. **d,** Comparison between the experimental (black curve) and simulated (green curve) integrated Raman spectra, obtained by summing

up all the TEPL spectra in the Raman shift representation acquired at different excitation wavelengths. The blue curve shows the calculated Raman spectrum for the ground electronic state.

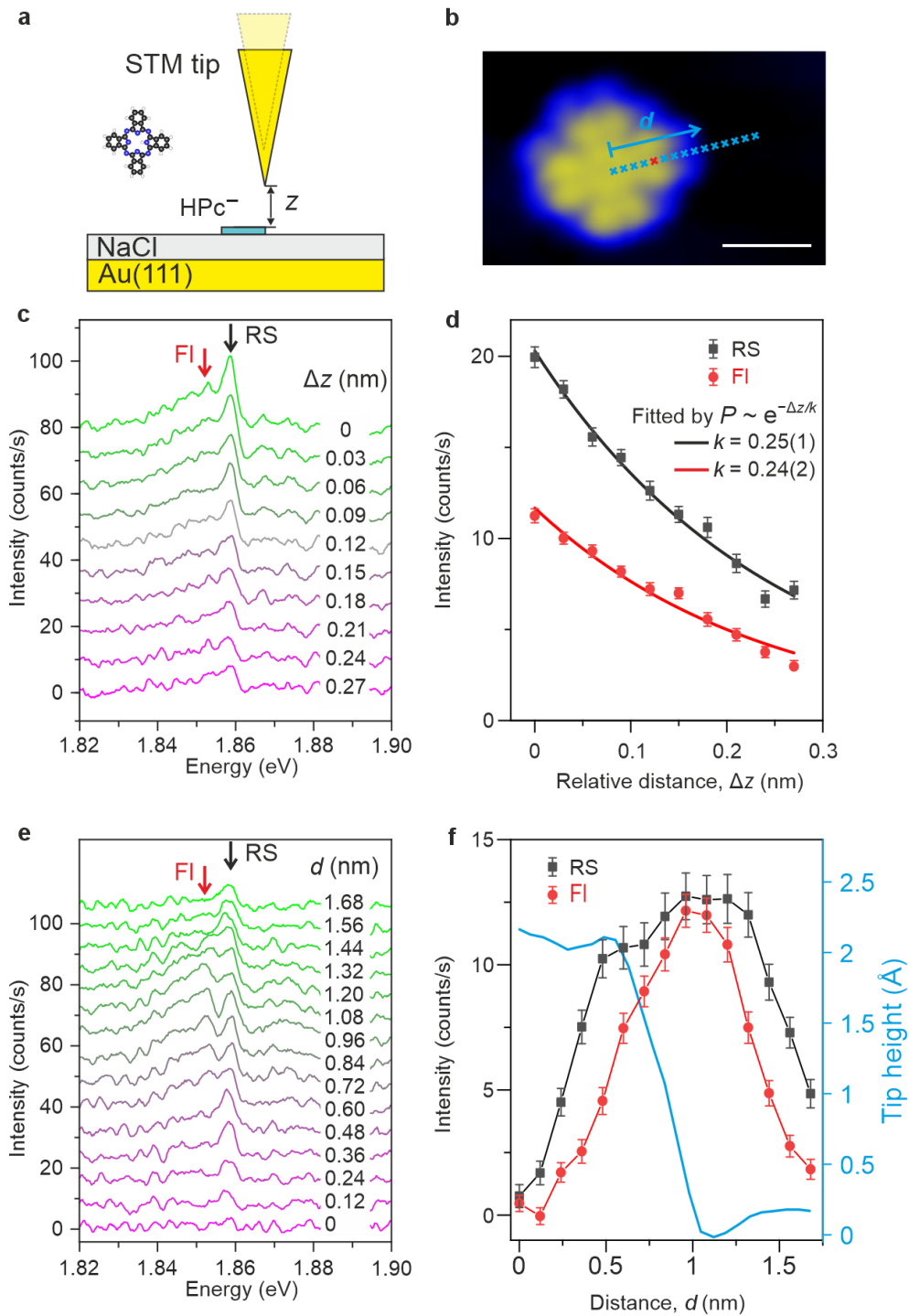

**Fig. 4: Ångström-scale resolution in selective excitation of molecular vibrations.** Schematic illustration for measuring the dependence of the resonance Raman scattering (RS) and fluorescence (Fl) signals upon

changing the plasmonic gap size **(a)** and the lateral position **(b)** of the nanotip over the HPc¯ molecule. The white scale bar indicates a length of 1 nm. **c,** TEPL spectra measured by retracting the nanotip with a step size of 30 pm over the molecular lobe (red cross in **b**). The STM junction was stabilized at $V = -1.5$ V, $I = 4$ pA before starting the measurement ($\Delta z = 0$ nm). **d**, Variation in the spectral intensities of the resonance Raman scattering (black dots, averaged over 1.856-1.861 eV) and fluorescence signals (red dots, averaged over 1.840-1.850 eV) upon relative change of the plasmonic gap size $\Delta z$. **e**, Molecular spectra recorded by placing the nanotip over the blue crosses shown in **b**, with a step size of 1.2 Å. **f**, Variation in the spectral intensities of the resonance Raman scattering (black dots, averaged over 1.856-1.861 eV) and fluorescence signals (red dots, averaged over 1.840-1.850 eV) upon change of the lateral position of the nanotip over the molecule. The STM was operated in the constant current mode at $V = -1.5$ V, $I = 4$ pA. The blue curve represents the relative height profile of the nanotip during the measurement. The spectra in **c** and **e** are vertically shifted for clarity. The laser wavelength was set at 633.5 nm, the laser power was kept constant at 0.23 mW, and the acquisition time was $t = 20$ s for all the spectra.

# Supplementary Information for

# Selective Excitation of Vibrations in a Single Molecule

Yang Luo[1,*], Shaoxiang Sheng[1], Michele Pisarra[2,3], Alberto Martin-Jimenez[1,4], Fernando Martin[4,5*], Klaus Kern[1,6], Manish Garg[1,*]

[1] Max Planck Institute for Solid State Research, Heisenbergstr. 1, 70569 Stuttgart, Germany

[2] Dipartimento di Fisica, Università della Calabria, Via P. Bucci, cubo 30C, 87036, Rende (CS), Italy

[3] INFN-LNF, Gruppo Collegato di Cosenza, Via P. Bucci, cubo 31C, 87036, Rende (CS), Italy

[4] Instituto Madrileño de Estudios Avanzados en Nanociencia (IMDEA Nano), Faraday 9, Cantoblanco, 28049 Madrid, Spain

[5] Departamento de Química, Módulo 13, Universidad Autónoma de Madrid, 28049 Madrid, Spain

[6] Institut de Physique, Ecole Polytechnique Fédérale de Lausanne, 1015 Lausanne, Switzerland

*Authors to whom correspondence should be addressed.

y.luo@fkf.mpg.de, fernando.martin@imdea.org and mgarg@fkf.mpg.de

**Contents**

# Section I. Fluorescence and Raman scattering processes in a molecule

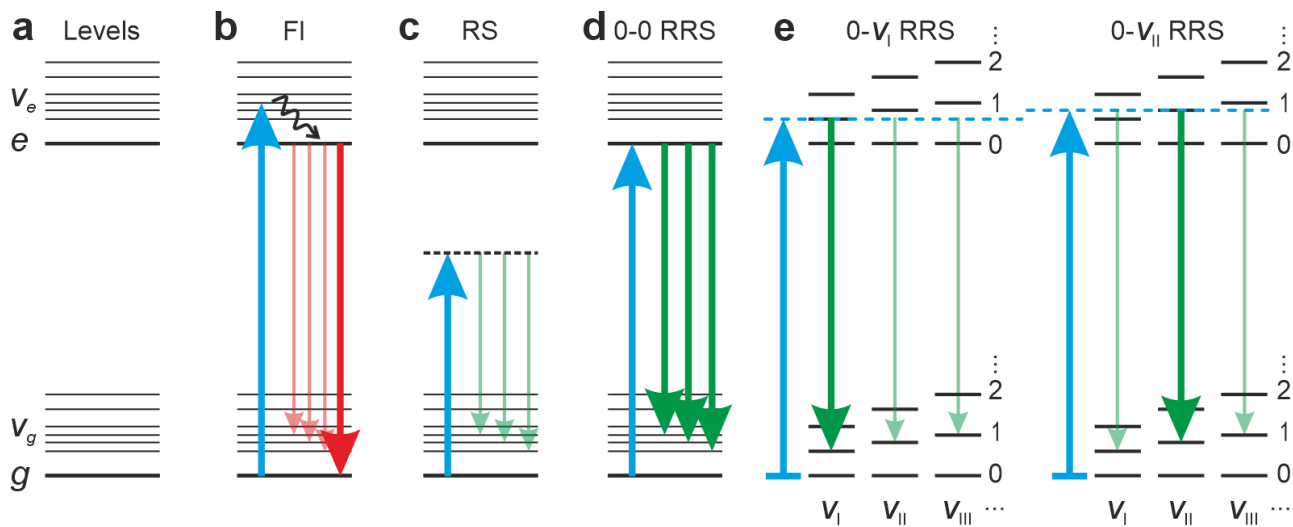

**Supplementary Fig. 1:** Schematic illustration of various processes ensuing photoexcitation in a molecule. **a,** The energy alignment of various vibrational modes associated with the ground (*g*) and excited (*e*) electronic states. **b, c, d,** Depiction of fluorescence (Fl), Raman scattering (RS), and resonance Raman scattering (RRS) processes in the molecule, respectively. **e,** Schematic description of vibrationally resolved resonance Raman scattering processes involving, the $v_\mathrm{I}$ (left panel) and $v_\mathrm{II}$ (right panel) vibrational modes of the ground electronic state, *g*. Numerals, 0, 1, 2… represent the quantum numbers of the various vibrational modes (e.g. $v_\mathrm{I}$ or $v_\mathrm{II}$) in the two electronic states. The molecule is initially in the vibrational ground state before the laser excitation.

Many processes ensue after the photoexcitation of a molecule. We start by considering the interaction of the light with the quantized electronic and vibrational levels of a single molecule, which is simplified as the energy levels as shown in Supplementary Fig. 1a. The electronic ground and the excited states are marked by *g* and *e*, respectively. The vibrational levels in the two electronic states are denoted by $v_g$ and $v_e$. Here, three light interaction processes are considered: fluorescence (Fl), Raman scattering (RS), and the resonance Raman scattering (RRS).

In the fluorescence (Fl) process as depicted in Supplementary Fig. 1b, an incident photon excites the system to one of the $|v_e, e\rangle$ levels. The system can redistribute the excess vibrational energy among its various modes by internal conversion or by exchanging energy with its environment to reach the bottom of the excited electronic state, i.e., $|0, e\rangle$. This state can then generate the fluorescence emission by the radiative transition to the ground electronic state. Several emission pathways are possible as indicated by red downward arrows. If the geometry molecule is similar in the ground and excited electronic states, the most intense transition (thick arrow) happens between the $|0,e\rangle$ and $|0,g\rangle$ states.

In the Raman Scattering (RS) process, the incident light is inelastically scattered by interacting with a vibrational mode of the molecule. The process can be sketched with the incident photon absorbed to a virtual state and then re-emitted with a lower energy, as shown in Supplementary Fig. 1c. For simplicity we consider only Stokes processes, which are relevant for the current experiment. Multiple Raman scatterings are possible and the energy loss of the scattered light depends on the energy of the excited vibrational modes. The Raman scattering probability is usually very low, as symbolized by the thin downward green arrows in Supplementary Fig. 1c.

The Raman scattering cross-section can increase dramatically if the incident light is resonant with an electronic transition in the molecule. This phenomenon is depicted in Supplementary Fig. 1d, where the incident light is resonant with the electronic transition between the $|0,g\rangle$ and $|0,e\rangle$ states. This electronic transition is also called the 0-0 transition, because it involves the vibrational ground states of the *g* and *e* electronic states. In this case, all the Raman modes coupled to the excited state are enhanced, as symbolized by the thick downward green arrows.

The energy of the incident light can be tuned to be resonant with the $|0,g\rangle$ and $|v_\text{I},e\rangle$ states, as shown in the left panel of Supplementary Fig. 1e. While all the Raman scattering processes from $|v_\text{I},e\rangle$ towards $|v_\text{I},g\rangle$, $|v_\text{II},g\rangle$, $|v_\text{III},g\rangle$,... are possible (thin downward green arrows), the one towards the $|v_\text{I},g\rangle$ state will be dramatically enhanced (thick downward green arrow) because of the dominance of the Frank-Condon factor $\langle v_\text{I},e|v_\text{I},g\rangle$, provided that the geometry of the molecule in the electronic *e* and *g* states is not very different. A similar scenario involving the $|v_\text{II}\rangle$ vibrational mode is depicted in the right panel of Supplementary Fig. 1e, when the energy of the incident light is tuned to be resonant with the $|0,g\rangle$ and $|v_\text{II},e\rangle$ states. Due to the narrow Raman linewidths in the single-molecule spectrum, distinguishing the resonance Raman transitions from different vibronic transitions is possible, as sketched in Supplementary Fig. 1e. Hence, by modulating the photon energy of the incident light, it is possible to selectively populate a specific vibrational mode in the ground state. In the vibrationally resolved resonant Raman scattering processes, different laser excitation photon energies will result into different enhancement factors of the Raman scattering rates. In all the cases, the most prominent Raman scattering occurs between identical vibrational modes in the two electronic states (thick green arrows in Supplementary Fig. 1e), i.e. when their energy difference will be almost identical to that of the fluorescence transition (thick red arrow in Supplementary Fig. 1b).

## Section II. Density functional theory calculations

Atomistic calculations have been performed with the Gaussian 16 program package[1]. We carried out unrestricted all electron calculations using the B3LYP functional[2] and the 631G(d,p) basis function. The resonance Raman spectroscopy (RRS) calculations have been carried out taking into account the excited state involved in the transition as implemented in the Gaussian 16 package[3,4]. We have conducted the simulations for both the $H_2Pc$ and the $HP_C^-$ molecules, and the results for the $HP_C^-$ molecule are described here in detail.

We optimized the geometry of the molecule in the ground state, obtaining also the vibrational spectrum. The Raman intensities of the ground electronic state are also obtained during this step. The ground electronic state of the $HP_C^-$ molecule is characterized by a flat geometry, as shown in the left panel of Supplementary Fig. 2. The vertical electronic excitation spectrum and transition dipole moments at the ground state geometry were calculated within a TD-DFT approach obtaining up to the 20$^{th}$ transition (see Supplementary Table 1). This procedure allowed us to identify the electronic transitions with nonzero oscillator strength. The S1 and S2 excited states correspond to the experimentally electronic states named $Q_x$ and $Q_y$[5,6]. In our case, the relevant excited state is the singlet state S1 (the third overall excited state, highlighted in red in Supplementary Table 1), which contributes to the fluorescence peaks in Fig. 1b of the main text. A geometry optimization was then performed for the molecules in the excited electronic S1, using the ground state geometry as the starting point. We first found a transition state, with one imaginary frequency in the vibrational spectrum, characterized by a flat arrangement of the atoms (S1-I in Supplementary Fig. 2). A further optimization of the geometry, with adding little displacements to the S1-I configuration, leads to the equilibrium configuration in the excited state (S1-F in Supplementary Fig. 2), which is characterized by a bent geometry.

**Supplementary Table 1:** $HP_C^-$ vertical transitions at the ground electronic state geometry.

| Ex. State | Ex. State # | Spin mult. | Energy (eV) | Energy (nm) | Oscillator Strength |
|---|---|---|---|---|---|
| T1 | 1 | 3 | 1.2364 | 1002.8 | 0 |
| T2 | 2 | 3 | 1.2926 | 959.22 | 0 |
| **S1** | **3** | **1** | **2.1365** | **580.3** | **0.3634** |
| S2 | 4 | 1 | 2.156 | 575.07 | 0.3677 |
| T3 | 5 | 3 | 2.414 | 513.61 | 0 |
| T4 | 6 | 3 | 2.4279 | 510.67 | 0 |
| T5 | 7 | 3 | 2.6629 | 465.6 | 0 |
| T6 | 8 | 3 | 2.7088 | 457.7 | 0 |
| S3 | 9 | 1 | 2.83 | 438.11 | 0 |
| S4 | 10 | 1 | 2.8359 | 437.2 | 0.0001 |
| T7 | 11 | 3 | 2.8755 | 431.17 | 0 |
| T8 | 12 | 3 | 2.8902 | 428.98 | 0 |
| T9 | 13 | 3 | 2.9519 | 420.02 | 0 |
| T10 | 14 | 3 | 3.003 | 412.87 | 0 |
| T11 | 15 | 3 | 3.0066 | 412.37 | 0 |
| T12 | 16 | 3 | 3.1949 | 388.06 | 0 |
| T13 | 17 | 3 | 3.2201 | 385.03 | 0 |
| S5 | 18 | 1 | 3.2202 | 385.02 | 0.1869 |
| T14 | 19 | 3 | 3.245 | 382.07 | 0 |
| S6 | 20 | 1 | 3.2965 | 376.1 | 0.1772 |

In the resonance Raman scattering calculation we took into consideration both the S1-F and the S1-I geometries as intermediate states. In the cases of the S1-F configuration, the overlap integral between the vibrational ground states gave very small values, due to the large differences in the geometries of the ground and the excited state. For this reason, we continued our analysis using the S1-I configurations, neglecting the first (imaginary) vibrational mode. The RRS calculation has been carried out within a time-independent approach, as implemented in the Gaussian software, using various incident light frequencies. Since the S1->S0 transitions are dipole-allowed, the calculation included the Frank-Condon terms only, while the second order Herzberg-Teller terms of the dipole matrix element expansion were not included. We assumed a 40 cm$^{-1}$ broadening for the incident light and a 10 cm$^{-1}$ broadening for the scattered Raman signal in the calculations.

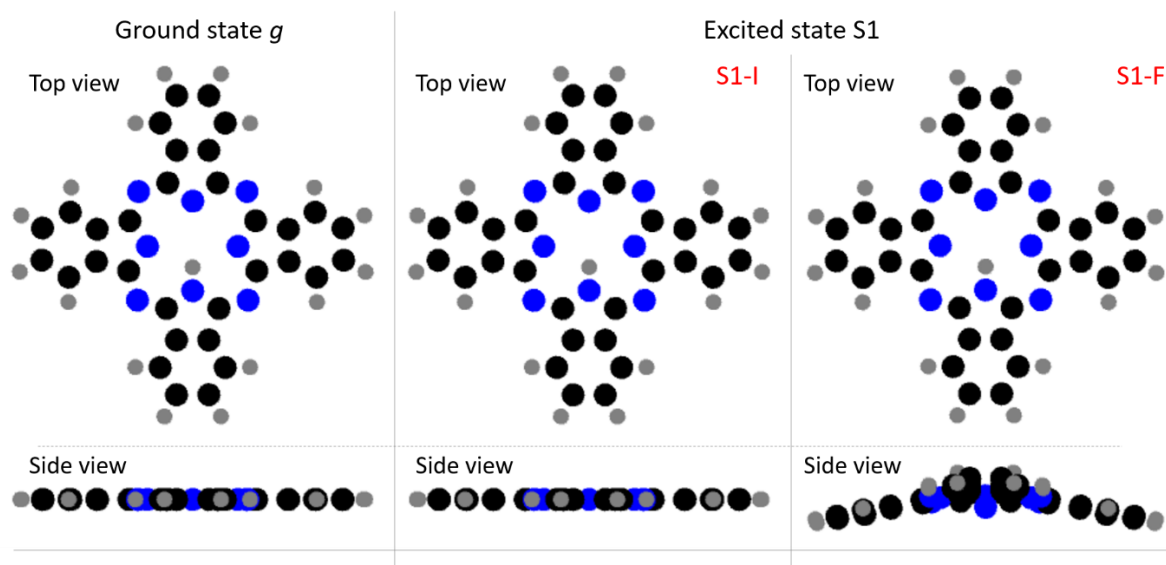

**Supplementary Fig. 2:** Geometry optimization of the HP$_C^-$ molecule in the ground state (*g*, left panel) and the S1 excited electronic state (central and right panels). The S1-I geometry corresponds to the atomic

configuration of the transition state, with one imaginary frequency; the S1-F geometry shows the equilibrium configuration in the excited state. The C, N, and H atoms are rendered as black, Blue and Gray disks, respectively.

Supplementary Fig. 3 shows the calculated resonance Raman scattering spectra for different photon energies of the incident light. To ease comparison with Fig. 2a of the main text, the spectra are plotted with their x-axis being the relative energy (in meV) with respect to the fluorescence emission peak in the range from −25 to 25 meV. Several peaks of different intensities can be clearly identified in the spectra. An apparent gradual shift of the narrow emission peaks is observed upon increasing the photon excitation energy of the laser, which is consistent with the experimental observations. The intensity of the Raman peaks is greatly enhanced when their transition energy approaches the energy of the 0-0 transition, whose spectral position is annotated by the vertical dashed red line in the plot.

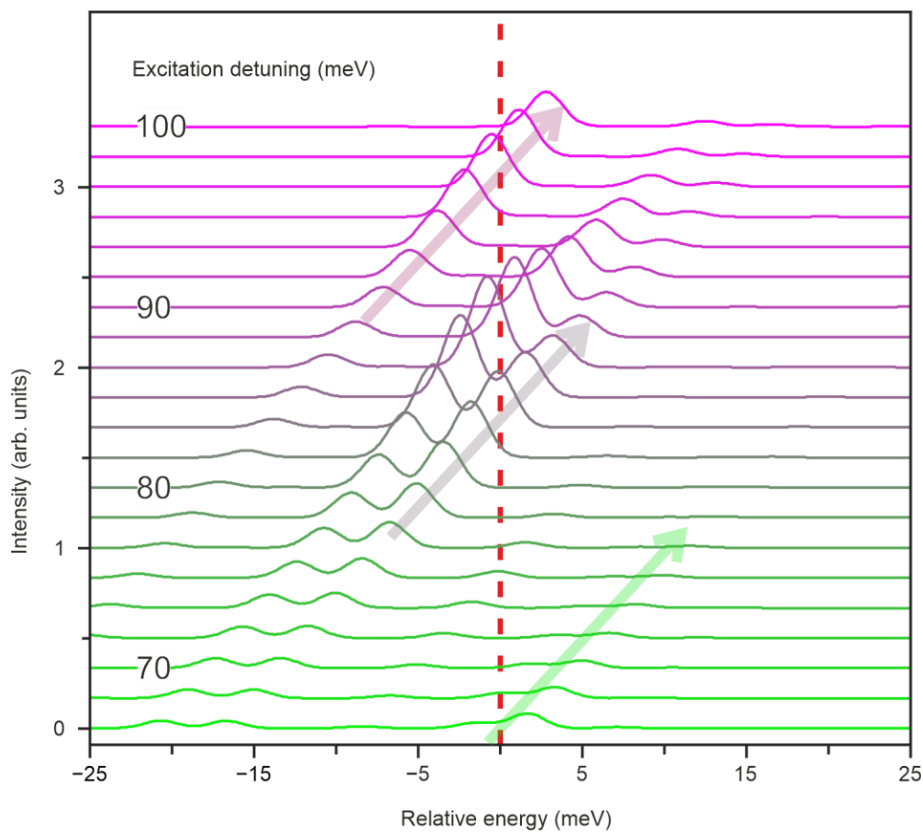

**Supplementary Fig. 3:** Calculated resonance Raman spectra for different photon energies of the incident light. The energy detuning between the incident photons and the calculated optical gap is annotated on the left side of the spectra. The resonance Raman spectra are plotted with their x-axis being the relative energy

(in meV) with respect to the 0-0 transition (fluorescence peak) at 616.05 nm (red dashed line). The spectra are displaced vertically for clarity.

**Section III. Optical setup for the single-molecule spectroscopy**

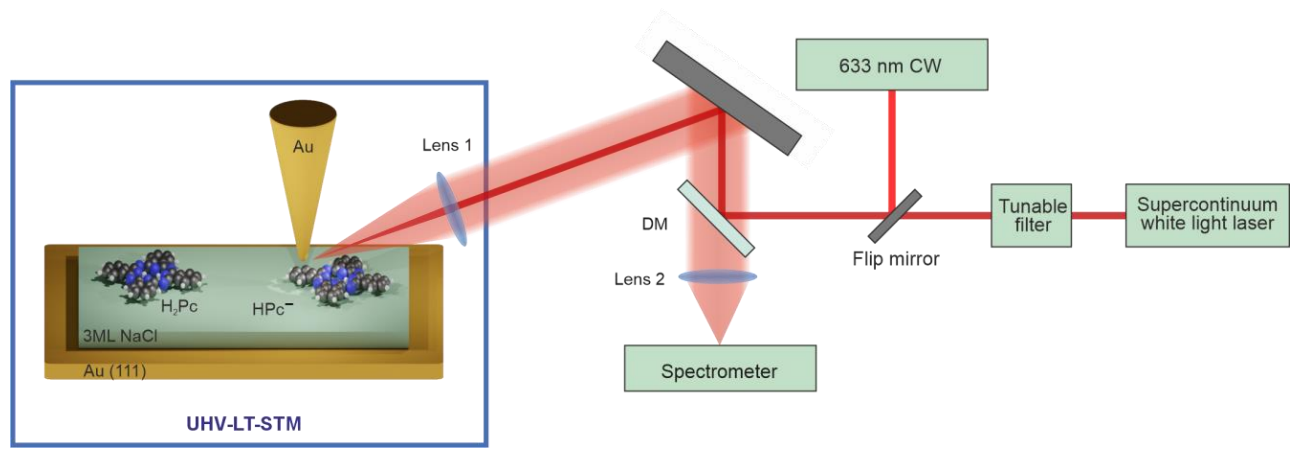

**Supplementary Fig. 4:** Optical setup for the single-molecule spectroscopy. Both the CW laser centered at 633 nm and the ps-laser with tunable excitation wavelength can be used as the excitation laser beam. A flip mirror is used to switch between the excitation sources of the CW laser and the ps-laser. DM: dichroic mirror.

**Section IV. Comparison between the experimental and the calculated Raman spectra**

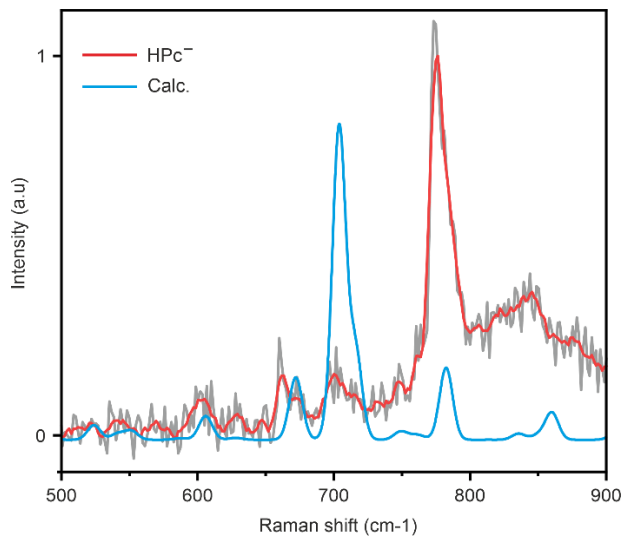

**Supplementary Fig. 5:** Comparison between the experimental TEPL spectrum from HPc⁻ molecule (red curve) and the calculated Raman spectrum for the ground electronic state (blue curve). The TEPL spectrum is adopted from Fig. 1 in the maintext, which was recorded with CW laser excitation ($\lambda \sim 633$ nm and laser power of 0.2 mW) at $V = -1$ V, $I = 4$ pA, and $t = 10$ s. The most prominent vibrational peak comes from the resonance Raman scattering, while the other vibrational modes are still visible in the tip-enhanced Raman spectroscopy.

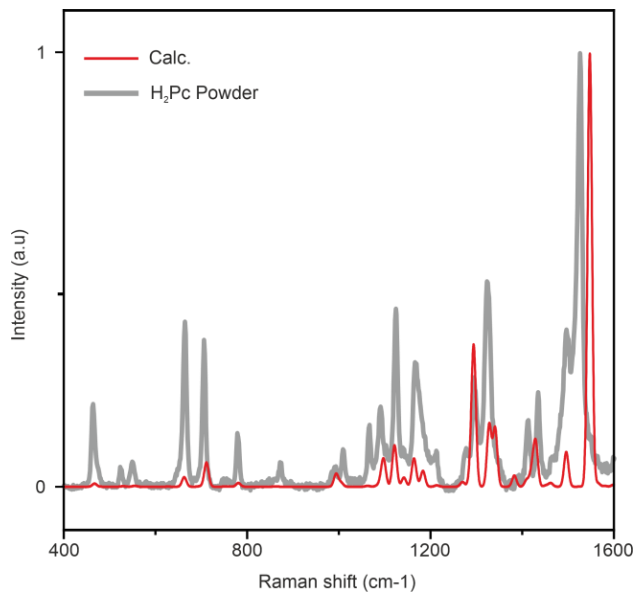

**Supplementary Fig. 6:** Comparison between the measured Raman spectrum from the powder of $H_2Pc$ molecules (gray curve) and the calculated Raman spectrum for the ground electronic state (red curve). The Raman spectra were recorded with CW laser excitation ($\lambda \sim 633$ nm and laser power of 0.2 mW).